\documentclass[aps,prl,twocolumn,superscriptaddress,groupedaddress,showpacs]{revtex4}

\usepackage[english]{babel}
\usepackage{dsfont} 

\usepackage{amsmath}
\usepackage{amsthm}
\usepackage{amssymb}
\usepackage{booktabs}
\usepackage[acronym]{glossaries}
\usepackage{hyperref}
\hypersetup{
  colorlinks   = true,  
  urlcolor     = blue,  
  linkcolor    = blue,  
  citecolor    = red    
}
\usepackage{natbib}
\usepackage{pgfplots}
\usepgfplotslibrary{groupplots}
\usepackage{siunitx}
\usepackage{subfigure}
\usepackage{tikz}
\usetikzlibrary{arrows,automata,backgrounds,decorations,positioning,shapes,snakes}
\usepackage[colorinlistoftodos,textsize=scriptsize]{todonotes}
\usepackage{url}
\usepackage{xifthen}

\usepackage{dsfont} 

\renewcommand{\d}[1]{\ensuremath{\operatorname{d}\!{#1}}}
\newcommand{\e}[1]{ {\mathrm{e}}^{ #1 } }

\newcommand{\expectation}[1]{ \mathbb{E} [ #1 ] }
\newcommand{\expectationBig}[1]{ \mathbb{E} \Bigl[ #1 \Bigr] }

\newcommand{\probability}[1]{ \mathbb{P} [ #1 ] }

\newcommand{\variance}[1]{ \mathrm{Var} [ #1 ] }
\newcommand{\varianceBig}[1]{ \mathrm{Var} \Bigl[ #1 \Bigr] }

\newcommand{\realNumbers}{ \mathbb{R} }

\newcommand{\refFigure}[1]{{\textrm{Figure~\ref{#1}}}}

\newcommand{\refEquation}[1]{{\textrm{\eqref{#1}}}}

\theoremstyle{definition}

\theoremstyle{remark}

\makeglossaries
\newacronym{ER}{ER}{Erd\"{o}s--R\'{e}nyi}

\begin{document}

\title{Sub-Poissonian Statistics of Jamming Limits in Ultracold Rydberg Gases}

\author{Jaron \surname{Sanders}}
\email{jaron.sanders@tue.nl}
\affiliation{Eindhoven University of Technology, P.O.\ Box 513, 5600MB Eindhoven, The Netherlands}

\author{Matthieu \surname{Jonckheere}}
\affiliation{University of Buenos Aires, Postal Code 1428, Buenos Aires, Argentina}
\email{mjonckhe@dm.uba.ar}

\author{Servaas \surname{Kokkelmans}}
\affiliation{Eindhoven University of Technology, P.O.\ Box 513, 5600MB Eindhoven, The Netherlands}
\email{s.kokkelmans@tue.nl}



\begin{abstract}
Several recent experiments 
have established by measuring the Mandel Q parameter 
that the number of Rydberg excitations in ultracold gases exhibits sub-Poissonian statistics. This effect is attributed to the Rydberg blockade that occurs due to the strong interatomic interactions between highly-excited atoms. Because of this blockade effect, the system can end up in a state in which all particles are either excited or blocked: a \emph{jamming limit}. 
We analyze appropriately constructed random-graph models that capture the blockade effect, and derive formulae for the mean and variance of the number of Rydberg excitations in jamming limits. This yields an explicit relationship between the Mandel Q parameter and the blockade effect, and comparison to measurement data 
shows strong agreement between theory and experiment.
\end{abstract}

\maketitle


Ultracold gases with atoms in highly excited states have attracted substantial interest over recent years, for example for their potential application in quantum computing \cite{lukin_dipole_2001,comparat_dipole_2010,jaksch_fast_2000}, and for study of non-equilibrium phase transitions \cite{weimer_quantum_2008}. These atomic systems exhibit complicated spatial behavior due to strong van der Waals or dipolar interactions between neighboring atoms, which has been demonstrated through several experimental observations of reduced fluctuation in the number of excitations in ultracold gases of Rydberg atoms \cite{liebisch_atom_2005,viteau_cooperative_2012,hofmann_sub-poissonian_2013,malossi_full_2014,schempp_full_2014,schauss_crystallization_2015}.

In these experiments, a laser facilitates excitation of ultracold atoms into a Rydberg state. After some time $t$, information on the mean and variance of the number of excited particles $X(t)$ is obtained by repeating counting experiments, and the Mandel Q parameter \cite{mandel_sub-poissonian_1979}
\begin{equation}
Q(t) = \frac{ \variance{X(t)} }{ \expectation{X(t)} } - 1
\label{eqn:Definition_of_the_Mandel_Q_parameter}
\end{equation}
is calculated to quantify a deviation from Poisson statistics, since if $X(t)$ is Poisson distributed, $Q(t) = 0$. The experiments establish that $X(t)$ is underdispersed, i.e.\ $Q(t) < 0$ for $t >0$, and $X(t)$ is said to have a sub-Poisson distribution.

The observed negative Mandel Q parameter is attributed to the Rydberg blockade effect \cite{lukin_dipole_2001,comparat_dipole_2010}. There exist simulation techniques \cite{petrosyan_two-dimensional_2013} and models based on Dicke states \cite{viteau_cooperative_2012} that numerically describe the Mandel Q parameter, but to the best of our knowledge, no closed-form expression is available in the literature that describes the relation between the Mandel Q parameter and the blockade effect.

Explicit formulae for the Mandel Q parameter are difficult to obtain, because the problem at hand is reminiscent of continuum random sequential adsorption problems \cite{evans_random_1993}. The standard two-dimensional continuum random sequential adsorption problem is that of throwing disks of radius $r > 0$ one by one randomly in a two-dimensional box, such that the disks do not overlap. This process continues until a jammed state is reached. Except for the one-dimensional variant, such problems are notoriously challenging to analyze due to spatial correlations.
One further question is whether such stochastic processes are suited to explain effects occurring in ultracold Rydberg gases, and if so, under what conditions. This matter is discussed in \cite{sanders_wireless_2014}, where a suitable stochastic process is provided based on rate equations that adequately describe the Rydberg gas when an incoherent process (such as spontaneous emission) occurs \cite{ates_many-body_2007}. 

This Letter adopts the stochastic process in \cite{sanders_wireless_2014} that models the Rydberg gas, and uses it to study the Mandel Q parameter in the jamming limit which occurs when atoms only transition from the ground state to the Rydberg state. The model includes the blockade effect through so-called interference graphs, and by considering specially constructed large \gls{ER} random graphs \cite{erdos_random_1959} that retain essential features of the blockade effect, we overcome the mathematical difficulties normally involved with having a spatial component. The problem remains nontrivial though, and we point interested readers to our rigorous derivation of the necessary fluid and diffusion limits \cite{bermolen_scaling_2015}. This Letter explains how to use our theoretical insights in the context of Rydberg gases through less complicated heuristic arguments, and while doing so explicitly relates the mean and variance of the number of excitations to the blockade effect. 




We consider a gas of ultracold atoms in an excitation volume $V \subseteq \realNumbers^3$, and we assume that each particle has its own distinct position. Each particle can go from a ground state to a Rydberg state, and a particle in the Rydberg state prevents neighboring particles from also entering the Rydberg state. The density of particles is assumed to be $\rho$, and the number of excitable particles $N$ within any region $A \subset V$ to be Poisson distributed with parameter $\rho A$. This implies in particular that in the absence of blockade effects, the number of excited particles within the excitation volume, $X(t)$, will be Poisson distributed, as is the case in experiments \cite{liebisch_atom_2005,viteau_cooperative_2012,hofmann_sub-poissonian_2013,malossi_full_2014,schempp_full_2014,schauss_crystallization_2015}. It also implies that the particles are uniformly distributed at random over the excitation volume. 

The blockade effect will be modelled using the notion of a blockade radius $r$. This is in line with simulations and measurements of pair correlation functions between atoms in the Rydberg state, which show a sharp cutoff when plotted as a function of the distance between the atoms \cite{robicheaux_many-body_2005,schauss_observation_2012}. Particles within a distance $r > 0$ are considered \emph{neighbors} of each other, and neighbors each block the other if excited. We denote the number of neighbors of a particle $i = 0, 1, \ldots, N$ within its blocking volume $V_{\mathrm{b},i}$ by $B_i$. As a consequence of our assumptions, the number of neighbors of particle $i$ is also Poisson distributed. Specifically,
\begin{align}
\probability{ B_i = b }
= \frac{ ( \rho V_{\mathrm{b},i} )^b \e{ - \rho V_{\mathrm{b},i} } }{ b! },
\quad b = 0, 1, \ldots,
\label{eqn:Number_of_neighbors_has_a_Poisson_distribution}
\end{align}
if $V_{\mathrm{b},i}$ is fully contained within $V$.

We will study the number of excitations by examining the asymptotic behavior of large \gls{ER} random graphs. Each vertex of such a graph will represent one particle, so the set of vertices is given by $\mathcal{V} = \{ 1, \ldots, N \}$. We draw an edge between two particles $i$ and $j$ if we consider particles $i$ and $j$ to be neighbors (particles that would block one another). One can construct an \gls{ER} random graph by considering every pair of vertices $(i,j)$ once, and drawing the edge between $i$ and $j$ with probability $p$, independent from all other edges.
In order to deduce information on $X(t)$ through examining the \gls{ER} random graph, we need to match the \gls{ER} random graph model to the physical system, and we will do so by counting and matching the number of neighbors. Matching the models has to be done via the number of neighbors, because there is \emph{no such notion as a physical position of a particle} in an \gls{ER} random graph. This principle, in fact, makes this mathematical model tractable.

\begin{figure}[!hbtp]
\begin{center}
\subfigure{
\includegraphics[width=0.57\linewidth]{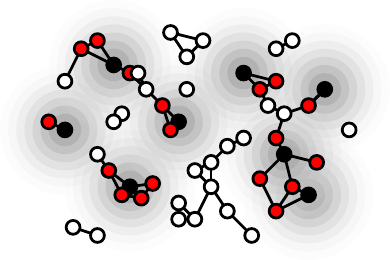} 
}
\subfigure{
\includegraphics[width=0.33\linewidth]{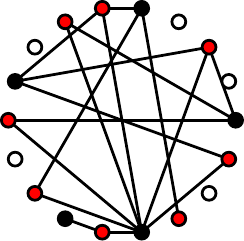} 
}
\caption{(left) A spatial Poisson point process in which neighbors within radius $r$ block each other is used to choose appropriate parameters for (right) an \gls{ER} random graph so that the particles have the same distribution of number of neighbors as $N \to \infty$.}
\label{fig:Interference_graph_and_ER_graph}
\end{center}
\end{figure}

The number of neighbors $B_{\mathrm{ER},i}$ of a particle $i$ in the \gls{ER} random graph is binomially distributed, $B_{\mathrm{ER}} \sim \mathrm{Bin}( N - 1, p )$, so that for $b = 0, 1, \ldots, N-1$,
$
\probability{ B_{\mathrm{ER},i} = b }
= \binom{ N-1 }{ b } p^b (1-p)^{N-1-b}
$,
and $\expectation{ B_{\mathrm{ER},i} } = (N-1)p$. When setting $p = c / N$ where $c$ is some constant, we see that as $N \to \infty$, the distribution converges to a Poisson distribution, 
\begin{equation}
\lim_{N \rightarrow \infty} \probability{ B_{\mathrm{ER},i} = b }
=  \frac{ c^b \e{-c} }{b!},
\quad b = 0, 1, \ldots. \label{eqn:Number_of_neighbors_in_an_ER_random_graph}
\end{equation}
Comparing \refEquation{eqn:Number_of_neighbors_in_an_ER_random_graph} to \refEquation{eqn:Number_of_neighbors_has_a_Poisson_distribution}, we note that the limiting distribution is the same if the average number of neighbors in the \gls{ER} random graph, $c$, is related to the density and blockade volume as
$
c = \rho V_{\mathrm{b}}
$.
By setting $c = \rho V_{\mathrm{b}}$, we ensure that the particles in the \gls{ER} random graph have the same distribution of number of neighbors as in the spatial problem when the number of particles $N \to \infty$. \refFigure{fig:Interference_graph_and_ER_graph} summarizes our construction.

\begin{figure}[!hbtp]
\begin{center}
\subfigure{
\includegraphics[width=0.30\linewidth]{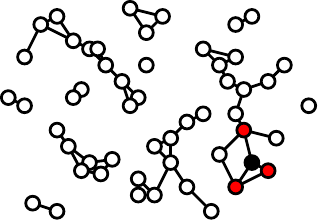} 
}
\subfigure{
\includegraphics[width=0.30\linewidth]{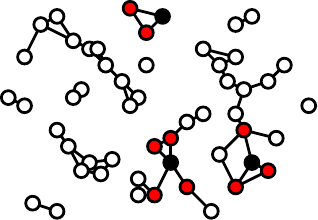} 
}
\subfigure{
\includegraphics[width=0.30\linewidth]{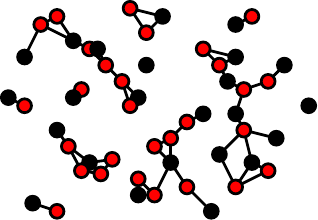} 
}
\caption{(left) A random first particle excites. (middle) Subsequently, random second and third particles excite. (right) The process continues until all particles are either blocked or excited, and the resulting state is a jamming limit.}
\label{fig:Dynamics_illustrated_of_a_jamming_limit}
\end{center}
\end{figure}

Let us now describe the dynamics, illustrated in \refFigure{fig:Dynamics_illustrated_of_a_jamming_limit}. At time $T_0 = 0$ the laser is activated, and from that point onward excitations can occur. At a time $T_1 > T_0$, the first particle (say $1$) excites and enters the Rydberg state. Due to the Rydberg blockade, particle $1$ will subsequently prevent all other particles within a radius $r$ from also exciting. Later, at a time $T_2 > T_1$, a second particle excites (say $2$), which cannot be within distance $r$ of particle $1$. Particle $2$ from that point onward also blocks particles within a distance $r$ of itself. This process continues until some finite time $T_{X(\infty)} < \infty$ when all particles are either blocked or excited. The random number of excited particles $1 \leq X(\infty) \leq N$ is then detected. 

We will now derive expressions for the Mandel Q parameter. 
Let $\mathcal{X}_m$ denote the excited particles at time $T_m$, and let $\mathcal{U}_m$ be the unaffected particles at time $T_m$. 
At time $T_0 = 0$, no vertices are excited or blocked, so $\mathcal{X}_0 = \emptyset$ and $\mathcal{U}_0 = \mathcal{V}$. At time $T_{m+1}$, a uniformly randomly chosen particle $v_{m+1} \in \mathcal{U}_m$ excites, and starts blocking random neighbors that were thus far unaffected, say $u_{m+1,1}, \ldots, u_{m+1,b} \in \mathcal{U}_m$, so that
$
\mathcal{X}_{m+1} = \mathcal{X}_m \cup \{ v_{m+1} \}
$, and
$
\mathcal{U}_{m+1} = \mathcal{U}_m \backslash ( \{ v_{m+1} \} \cup  \{ u_{m+1,1}, \ldots, u_{m+1,b} \} )
$.
This stochastic process continues until the moment $\tau$ a jamming limit occurs, i.e.\ when $\mathcal{U}_{\tau} = \emptyset$ and the number of unaffected particles equals zero. 

The number of unaffected particles $U_m = | \mathcal{U}_m |$ can be described using a stochastic recursion. When the $(m+1)$-th excitation occurs, the number of unaffected particles decreases by (i) the one particle that excites, and (ii) a random number of unaffected particles that each is neighbor of the new excitation with probability $p$ and thus now become blocked. Conditional on there being $N = n$ particles in the excitation volume, we have
\begin{equation}
U_{m+1}
= U_m - 1 - \mathrm{Bin}( U_m - 1, p ),
\quad
U_0 = n.
\label{eqn:Stochastic_recursion_for_Um}
\end{equation}
We will now analyze the stochastic recursion in \refEquation{eqn:Stochastic_recursion_for_Um}, and identify the moment $\tau$ the number of unaffected particles is zero, i.e.\ $U_\tau = 0$. Precisely at this moment, we have that the number of excitations $X(\infty) = \tau$. 

Our supplemental material details the following steps \cite{supplementary_material_of_jamming_limits}. From \refEquation{eqn:Stochastic_recursion_for_Um}, we obtain a closed-form expression for $\expectation{U_m}$ by invoking the tower property and giving an induction argument. Through decomposition, we subsequently obtain an expression for $\variance{U_m}$. 
When scaling the probability of being neighbors as $p = c / n$, the mean and variance converge to fluid limits, which can be seen by letting $f \in [0,1]$, and proving that as $n \to \infty$,
\begin{equation}
\frac{ \expectation{ U_{[fn]} } }{n} 
\to u(f),
\quad
\frac{ \variance{ U_{[fn]} } }{n}
\to v(f).
\end{equation}
Here, $[\cdot]$ denotes rounding to the nearest integer, and the fluid limits are $u(f) = \e{-cf} - ( 1 - \e{-cf} ) / c$, and $v(f) = ( e^{-c f} ( 1 - e^{-c f} ) ( (1+2c)e^{-c f} - 1 ) ) / (2 c)$. Note that these fluid limits are rigorously proven in \cite{bermolen_scaling_2015}.

\begin{figure}[!hbtp]
\begin{center}
\small
\includegraphics[width=\columnwidth]{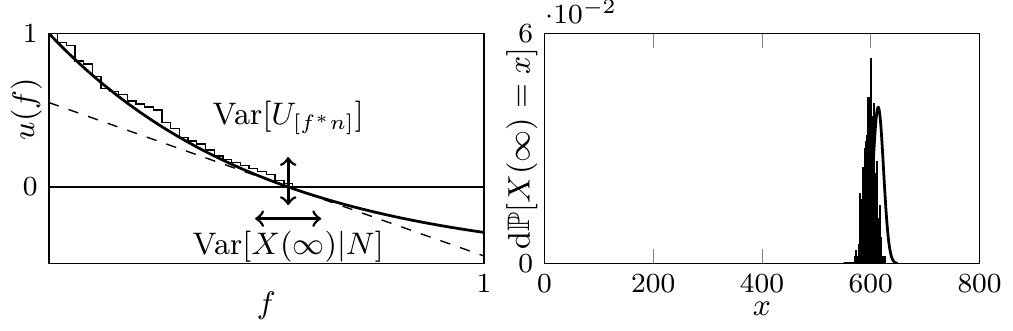}
\end{center}
\vspace{-1em}
\caption{\textrm{(left) The fluid limit $u(f)$ together with a sample path of $U_m / n$ for $n=50$. The dashed line indicates the tangent at $f^*$, and the arrows indicate the typical fluctuations. (right) Histogram of the number of excitations in a two-dimensional random sequential adsorption problem, and precisely $n = 800$ particles, together with the probability density function of a normal distribution with mean $n \ln{(1+c)} / c$ and variance $n c/(2(1+c)^2)$.}}
\label{fig:Figure__Explaining_the_derivative_trick__Histogram_of_Poisson_disk_throwing_processes}
\end{figure}

Consider now \refFigure{fig:Figure__Explaining_the_derivative_trick__Histogram_of_Poisson_disk_throwing_processes}~(left) and the following steps. The process $U_{m}$ hits zero when $m \approx f^* n$, with $f^* = \ln{(1+c)/c}$ being the solution to $u(f^*) = 0$. 
Therefore,
\begin{equation}
\expectation{X(\infty) | N} 
\approx f^* n
= \frac{ n \ln{(1+c)} }{c}.
\label{eqn:Conditional_mean_of_Xinfty}
\end{equation}
To approximate the variance, calculate $u'(f)$ and note that $u'(f^*+\varepsilon) \approx -1$ for sufficiently small $\varepsilon$. Since $U_m$ is probably near $0$ for $m \approx f^*n$, the fluctuations in $X(\infty)$ will thus be of the order of $\sqrt{ \variance{ U_{[f^*n]} } }$. Hence,
\begin{equation}
\variance{ X(\infty) | N } 
\approx \variance{U_{[f^*n]}} 
\approx v(f^*) n
= \frac{nc}{2(1+c)^2}.
\label{eqn:Conditional_variance_of_Xinfty}
\end{equation}
Invoking the central limit theorem, we have for large fixed $n$ that the number of excitations is approximately normal distributed with mean $n \ln{(1+c)}/c$ and standard deviation $\sqrt{ n c / (2(1+c)^2) }$. This, \refEquation{eqn:Conditional_mean_of_Xinfty}, and \refEquation{eqn:Conditional_variance_of_Xinfty} are formally established by deriving diffusion limits in \cite{bermolen_scaling_2015}.

%
%
%



Let us compare \refEquation{eqn:Conditional_mean_of_Xinfty} and \refEquation{eqn:Conditional_variance_of_Xinfty} to simulations of the mean and variance observed in the two-dimensional random sequential adsorption problem described earlier, and with periodic boundary conditions. We consider $h = 1 \si{\micro\metre}$, $l = w = 400 \si{\micro\metre}$, $r = 6.5 \si{\micro\metre}$, and $\rho = 5 \times 10^9 \si{\per\cubic\centi\metre}$, which are typical values in magneto-optical traps, and correspond to $n \approx 800$ and $c \approx 0.664$. \refFigure{fig:Figure__Explaining_the_derivative_trick__Histogram_of_Poisson_disk_throwing_processes}~(right) shows a histogram of the number of excitations, as well as the probability density function of a normal distribution with mean $n \ln{(1+c)} / c$ and variance $n c/(2(1+c)^2)$. Comparing to the simulation's outcome, our expressions differ for this set of parameters (i) $2.6\%$ for the mean, (ii) $2.5\%$ for the variance, and (iii) $0.015\%$ for the Mandel Q parameter. Because the mean and variance are both overestimated, the Mandel Q parameter happens to be more accurately approximated. The errors our approximation makes can be attributed to the fact that particles in the Erd\"{o}s-R\'{e}nyi random graph model have no physical position, whereas particles in two-dimensional Poisson disk throwing processes do exhibit spatial correlations. Intriguingly the random graph, which has no spatial interpretation, yields a good approximation.

It is important to understand that the results thus far are conditional on there being $N = n$ particles within the excitation volume. However, the number of particles $N \sim \mathrm{Poi}(\rho V)$ is itself random. To obtain an unconditional expression for the mean and variance, we can utilize the tower property,
$
\expectation{X(\infty)} 
= \expectation{ \expectation{ X(\infty) | N } }
\approx \expectation{ N } \ln{(1+c)} / c
= \rho V \ln{(1+c)}/c
$,
and decomposition,
$
\variance{X(\infty)}
= \expectation{ \variance{X(\infty)|N} } + \variance{ \expectation{X(\infty)|N} }
\approx \expectation{ N c / (2(1+c)^2) } + \variance{ N \ln{(1+c)}/c }
= ( c/(2(1+c)^2) + (\ln{(1+c)}/c)^2 ) \rho V
$.
Recalling definition \refEquation{eqn:Definition_of_the_Mandel_Q_parameter}, the Mandel Q parameter in the jamming limit is therefore
\begin{equation}
Q(\infty) 
\approx  \frac{c^2}{2(1+c)^2\ln{(1+c)}} + \frac{\ln{(1+c)}}{c} - 1,
\label{eqn:Unconditional_Mandel_Q_parameter}
\end{equation}
which is exact in the \gls{ER} case when $\rho V \to \infty$ \cite{bermolen_scaling_2015}. Note that \refEquation{eqn:Unconditional_Mandel_Q_parameter} only depends on the average number of neighbors $c$, which in fact explains observations on simulated Mandel Q parameters \cite{petrosyan_two-dimensional_2013} as we discuss in \cite{supplementary_material_of_jamming_limits}. 

%
%
%


Let us also discuss the time-dependency of the mean number of excitations. We incorporate time-dependency by assuming that every unaffected particle excites at rate $\lambda$, and specifically that $T_m - T_{m-1} \sim \mathrm{Exp}(\lambda U_{m-1})$, which corresponds to modelling the Rydberg gas using rate equations \cite{ates_many-body_2007} as discussed in \cite{sanders_wireless_2014}.
Under these assumptions, 
we 
obtain the time-dependent fluid limit \cite{bermolen_scaling_2015}	
\begin{equation}
\frac{ \expectation{ X(t) | N } }{n} 
\to x(t) 
= \lambda \int_0^t u(x(s)) ds.
\label{eqn:Fluid_limit_for_Xt}
\end{equation}
After substituting $u(f) = \e{-cf} - ( 1 - \e{-cf} ) / c$ into \refEquation{eqn:Fluid_limit_for_Xt}, recalling that initially no particles are excited, and taking the derivative, we obtain the differential system
$
\d{}x / \d{t}
= \lambda ( \exp{(-c x(t))} - ( 1 - \exp{(-c x(t))} ) / c )
$, with $x(0) = 0$,
for $x(t)$. This differential system has as its unique solution
$
x(t) 
= \ln{ ( 1 + c - c \e{-\lambda t} } ) / c
$,
and in particular, we recover the mean fraction of excitations in the jamming limit by calculating $\lim_{t \to \infty} x(t) = \ln{(1+c)} / c$. 

We now validate our model by comparisons with experimental data in \cite{viteau_cooperative_2012,hofmann_sub-poissonian_2013}, which requires us to incorporate the notion of a detector efficiency $\eta \in [0,1]$ into the model. The detector efficiency $\eta$ can be interpreted as being the probability that a Rydberg atom is detected. Let $I_i \sim \mathrm{Ber}(\eta)$ denote random variables that indicate whether each $i$-th Rydberg atom is detected. The number of detected Rydberg atoms is then given by $X_D(t) = \sum_{i=1}^{X(t)} I_i$. Assuming the $I_1, \ldots, I_{X(t)}$ are independent, calculation shows that
$
\expectation{X_D(t)} 
= \eta \expectation{ X(t) }
$,
and 
$
\variance{X_D(t)}
= \eta^2 \variance{ X(t) } + \eta(1-\eta) \expectation{ X(t) }
$,
see our supplementary material \cite{supplementary_material_of_jamming_limits}. The detected Mandel Q parameter thus reduces to 
$
Q_D(t) 
= \eta Q(t)
$, see also \cite{liebisch_atom_2005}.

\begin{figure}[!hbtp]
\begin{center}
\small
\includegraphics[width=0.9\linewidth]{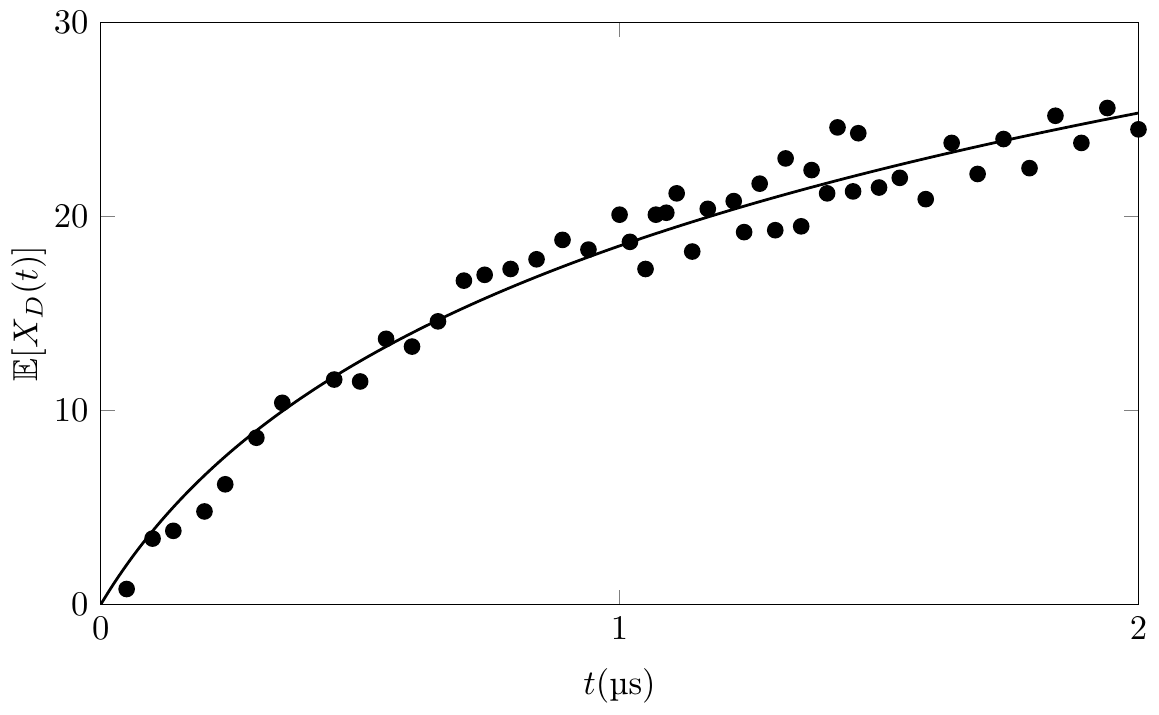}
\end{center}
\vspace{-2em}
\caption{\textrm{The average number of detected excitations as a function of time, $\expectation{X_D(t)}$, fitted to the measurement data in \cite[Fig.~1(a)]{viteau_cooperative_2012}. The fit results in an excitation rate of $\lambda = 14 \si{\kilo\hertz}$, and average number of neighbors of $c = 2.7 \times 10^2$.}}
\label{fig:Figure__Comparison_to_Viteau_2012}
\end{figure}

The experiments in \cite{viteau_cooperative_2012} were on excitation volumes said to contain $\rho V = 8 \times 10^3$ ground-state atoms, and with a reported detector efficiency of $\eta = 0.40$. Fitting
\begin{equation}
\expectation{ X_{\mathrm{D}}(t) } 
\approx \frac{\eta \rho V \ln{ ( 1 + c - c \e{-\lambda t} ) } }{c}
\end{equation} 
to measurements of the number of excitations as a function of time \cite[Fig.~1(a)]{viteau_cooperative_2012}, we obtain an excitation rate of $\lambda = 14 \si{\kilo\hertz}$, and average number of neighbors of $c = 2.7 \times 10^2$. \refFigure{fig:Figure__Comparison_to_Viteau_2012} shows strong agreement between theory and experiment.

\begin{figure}[!hbtp]
\begin{center}
\small
\includegraphics[width=0.9\linewidth]{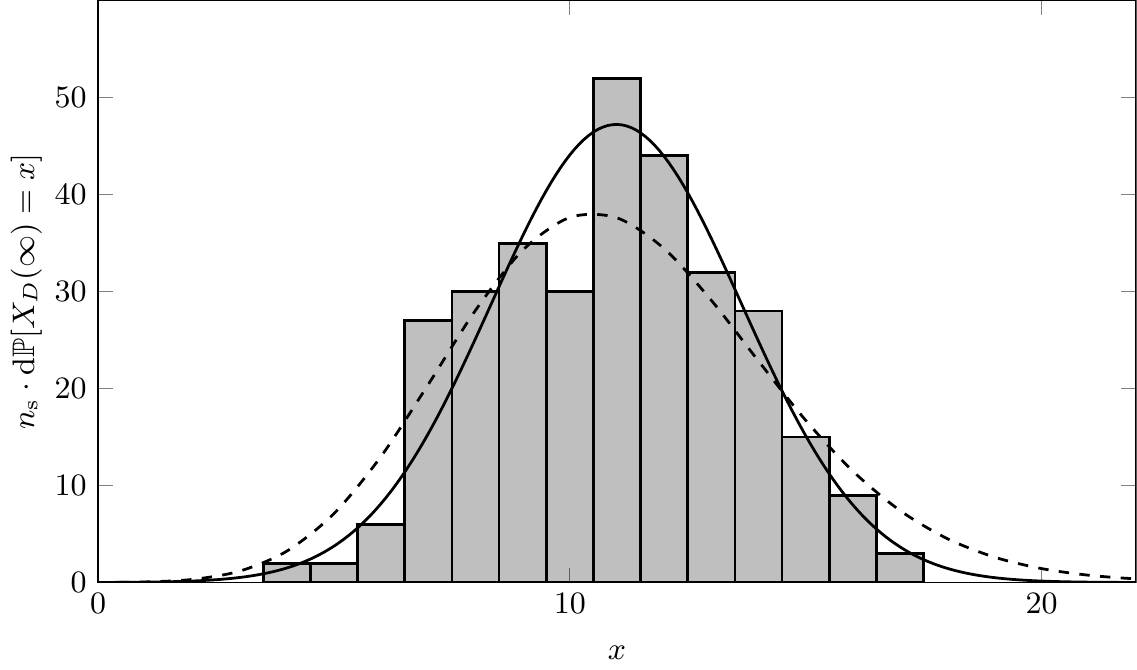}
\end{center}
\vspace{-2em}
\caption{\textrm{Histogram \cite[Fig.~4(a)]{hofmann_sub-poissonian_2013} of the number of detected Rydberg atoms, together with the appropriately scaled probability density function of a normal distribution with mean $\expectation{X_D(\infty)}$ and variance $\variance{X_D(\infty)}$. Here, $Q_D(\infty) \approx -0.36$, and the dashed line indicates the Poisson distribution.}}
\label{fig:Figure__Comparison_to_Weidemuller_2013}
\end{figure}

Lastly, we will compare our model to a histogram of the number of detected dark-state polaritons in \cite{hofmann_sub-poissonian_2013}. The histogram displays sub-Poissonian statistics due to a blockade effect that is a result of the dominant Rydberg character of the polaritons. Because of a partial overlap between the excitation laser and the cigar-shaped atomic cloud, we will infer the size of the excitation volume using the density $\rho = 5 \times 10^{17} \si{\per\cubic\meter}$ \cite{hofmann_sub-poissonian_2013} as follows. The detector efficiency is reported to be $\eta = 0.4$, and the histogram has a sample mean of $\expectation{X_D(\infty)} \approx 11$. If we assume that the blockade regions are spherical, and since the blockade radius $r \approx 5 \si{\micro\meter}$ \cite{hofmann_sub-poissonian_2013}, we find that $c = \tfrac{4}{3} \rho \pi r^3 \approx 2.6 \times 10^2$. Using our formula for the mean number of detected Rydberg atoms, it follows that $V = c \expectation{X_D(\infty)} / ( \rho \eta \ln{ (1+c) } ) \approx 2.6 \times 10^{-15} \si{\cubic\meter}$. 
The factor with which the density function of the Poisson distribution is scaled in \cite[Fig.~4(a)]{hofmann_sub-poissonian_2013} is $n_{\mathrm{s}} \approx 315$. 
\refFigure{fig:Figure__Comparison_to_Weidemuller_2013} now compares the appropriately scaled probability density function of a normal distribution with mean and variance as predicted by the model to the histogram in \cite[Fig.~4(a)]{hofmann_sub-poissonian_2013}. Our result $Q_D \approx -0.36$ is consistent with their observation that $Q_D = -0.32 \pm 0.04$ in the density range $2 \times 10^{17} \si{\per\cubic\meter} < \rho < 2 \times 10^{18} \si{\per\cubic\meter}$.

This Letter derived closed-form expressions for the Mandel Q parameter in limiting large random graphs constructed to model the spatial problem. This approach allowed us to derive explicit formulae for the mean and variance of the number of Rydberg excitations in the jamming limit, that turn out to be functions only of the average number of neighbors within the blockade volume. Our comparison to measurement data of \cite{viteau_cooperative_2012,hofmann_sub-poissonian_2013} shows quantitative agreement between theory and experiment, and we conclude that the model captures blockade effects observed in ultracold Rydberg gases.

Interesting future research would be to explore the approximating relation between random graphs and spatial problems, particularly because higher-dimensional continuum random sequential adsorption processes are difficult to analyse. The underlying stochastic recursions can also be generalized to incorporate additional effects \cite{bermolen_scaling_2015}, such as a slower reduction in the number of neighbors as more particles become blocked, and this can potentially extend the use of random graphs as an approximation to particle systems that exhibit complicated interactions.

\begin{acknowledgments}
This research was financially supported by an ERC Starting Grant, as well as The Netherlands Organization for Scientific Research (NWO), and is part of the research program of the Foundation for Fundamental Research on Matter (FOM). It was supported by a Wiskundecluster STAR visitor grant. We also acknowledge the European Union H2020 FET Proactive project RySQ (grant N.\ 640378). The authors are grateful for the support from Paola Bermolen, Sem Borst, Johan van Leeuwaarden, and Edgar Vredenbregt. 
\end{acknowledgments}

\bibliographystyle{apsrev}
\bibliography{G:/SOR/Bibliography/Bibliography}

\section{Supplementary material}

%

\subsection{Solving the stochastic recursions}

Our exposition led to the following stochastic recursions for the number of such particles,
\begin{gather}
X_{m+1} = m + 1,
\nonumber \\ 
U_{m+1} = U_m - 1 - \mathrm{Bin}( U_m - 1, p ),
\label{eqn:Supplementary_Material__Stochastic_recursions}
\end{gather}
with $X_0 = 0$ and $U_0 = n$. These recursions can be leveraged to determine the mean and the variance of the number of unaffected particles at each moment $m$ excitations have occurred, i.e.\ $\expectation{U_m}$ and $\variance{U_m}$. To see this, let $m \geq 1$ and start by noting that
\begin{equation}
U_{m} \overset{d}= \mathrm{Bin}( U_{m-1} - 1, 1 - p ).
\end{equation}
Utilizing the tower property, we find that
\begin{align}
\expectation{ U_{m} } 
&
= \expectation{ \expectation{ \mathrm{Bin}( U_{m-1} - 1, 1 - p ) | U_m } }
\nonumber \\ &
= (1-p) \expectation{ U_{m-1} } - (1-p).
\end{align}
Iterating and recalling that $\expectation{ U_0 } = n$, we obtain
\begin{align}
\expectation{ U_{m} }
& 
= (1-p)^{m} n - \sum_{i=1}^{m} (1-p)^i 
\nonumber \\ &
= (1-p)^{m} n - \frac{(1-p) - (1-p)^{m+1}}{p}.
\label{eqn:Supplementary_Material__Expression_for_the_expectation_of_Um}
\end{align}
A recursion for the variance can be found in a similar fashion by first decomposing
\begin{align}
&
\variance{U_m}
= \variance{ \expectation{ \mathrm{Bin}( U_{m-1} - 1, 1 - p ) | U_{m-1} } } 
\nonumber \\ &
\phantom{\variance{U_m} =} + \expectation{ \variance{ \mathrm{Bin}( U_{m-1} - 1, 1 - p ) | U_{m-1} } }
\nonumber \\ &
= \variance{ (U_{m-1}-1)(1-p) } + \expectation{ (U_{m-1}-1)p(1-p) }
\nonumber \\ &
= (1-p)^2 \variance{U_{m-1}} + p(1-p) ( \expectation{U_{m-1}} - 1 ),
\end{align}
and then recalling that $\variance{U_0} = 0$, $\expectation{U_0} = n$, so that after iterating,
\begin{align}
\variance{U_m} 
&
= p(1-p)^{2m-1} (n-1) 
\nonumber \\ &
\phantom{=} + p \sum_{i=1}^{m-1} (1-p)^{2i-1} ( \expectation{U_{m-i}} - 1 ).
\label{eqn:Supplementary_Material__Iterated_variance_of_Um}
\end{align}
Substituting \refEquation{eqn:Supplementary_Material__Expression_for_the_expectation_of_Um} into \refEquation{eqn:Supplementary_Material__Iterated_variance_of_Um} and simplifying, we find that
\begin{widetext}
\begin{equation}
\variance{U_m}
= \frac{(p-2) (1-p)^m ((n-1) p+1)+(1-p)^{2 m} (1-(n-1) (p-2) p)-p+1}{(p-2) p}.
\label{eqn:Supplementary_Material__Expression_for_the_variance_of_Um}
\end{equation}
\end{widetext}


\subsection{Determining fluid limits}

When $n \to \infty$ and $p = c / n$, there exist fluid limits for $\expectation{U_m}$ and $\variance{U_m}$. To see this, define 
\begin{equation}
u(f) = \lim_{n \to \infty} \expectation{ U_{[fn]} } / n
\end{equation}
for $f \in [0,1]$, with $[\cdot]$ denoting rounding to the nearest integer. Utilizing \refEquation{eqn:Supplementary_Material__Expression_for_the_expectation_of_Um}, we find that
\begin{align}
u(f) 
&
= \lim_{n \to \infty} \Bigl( 1 - \frac{c}{n} \Bigr)^{[fn]} - \frac{1}{c} \Bigl( 1 - \frac{c}{n} \Bigr) \Bigl( 1 - \Bigl( 1 - \frac{1}{c} \Bigr)^{[fn]} \Bigr)
\nonumber \\ &
= \e{-cf} - \frac{1}{c} ( 1 - \e{-cf} ).
\label{eqn:Supplementary_Material__Fluid_limit_of_the_expected_number_of_unaffected_particles}
\end{align}
Similarly, after defining 
\begin{equation}
v(f) = \lim_{n \to \infty} \variance{U_{[fn]}} / n,
\end{equation}
and substituting \refEquation{eqn:Supplementary_Material__Expression_for_the_variance_of_Um}, we find that
\begin{equation}
v(f) 
= \frac{ e^{-c f} ( 1 - e^{-c f} ) ( (1+2c)e^{-c f} - 1 ) }{2 c}.
\label{eqn:Supplementary_Material__Fluid_limit_of_the_variance_in_number_of_unaffected_particles}
\end{equation}

\subsection{Incorporating detector efficiency}

Assuming the $I_1, \ldots, I_{X(t)}$ are independent, the average number of detected Rydberg atoms is
\begin{align}
&
\expectation{X_D(t)} 
= \expectationBig{ \sum_{i=1}^{X(t)} I_i }
\nonumber \\ &
= \expectationBig{ \expectationBig{ \sum_{i=1}^{X(t)} I_i \big| X(t) } }
= \eta \expectation{ X(t) }.
\end{align}
The variance of the number of detected Rydberg atoms is 
\begin{align}
&
\variance{X_D(t)}
= \varianceBig{ \sum_{i=1}^{X(t)} I_i }
\nonumber \\ &
= \varianceBig{ \expectationBig{ \sum_{i=1}^{X(t)} I_i \big| X(t) } } + \expectationBig{ \varianceBig{ \sum_{i=1}^{X(t)} I_i \big| X(t) } }
\nonumber \\ &
= \eta^2 \variance{ X(t) } + \expectationBig{ \sum_{i=1}^{X(t)} \variance{ I_i } }
\nonumber \\ &
= \eta^2 \variance{ X(t) } + \eta(1-\eta) \expectation{ X(t) }.
\end{align}
The detected Mandel Q parameter is therefore
\begin{align}
&
Q_D(t) 
= \frac{ \variance{X_D(t)} }{ \expectation{X_D(t)} } - 1
\nonumber \\ &
= \eta \Bigl( \frac{ \variance{X(t)} }{ \expectation{X(t)} } - 1 \Bigr)
= \eta Q(t),
\end{align}
which completes the derivation.

\subsection{Comparison to Petrosyan, 2013}

Reference \cite{petrosyan_two-dimensional_2013} describes usage of semiclassical Monte Carlo simulations to study stationary states of the Rydberg gas in a two-dimensional system. There, particles are positioned on points of a lattice with spacing $a = 532 \si{\nano\meter}$, that fall within a circular excitation area with radius $R$, which is varied relative to the blockade radius of $r \approx 1.905 \si{\micro\meter}$. Ref.~\cite{petrosyan_two-dimensional_2013} finds numerically that $Q \approx -0.84$ for $R \gtrsim r$. This independence on the system size is explained by our model, because \refEquation{eqn:Conditional_mean_of_Xinfty} and \refEquation{eqn:Conditional_variance_of_Xinfty} indicate that the Mandel Q parameter only depends on the average number of neighbors $c$, which in this simulation setup approaches a constant for sufficiently large $R$. Modelling the blockade area as a hard circle of radius $r$, we have by Gauss's circle problem that $c+1 \approx 37$ lattice points fall within the blockade area for sufficiently large $R$. Because the number of particles within the excitation area $N$ did not fluctuate between simulation instances, we can use the conditional expressions for the mean and variance to estimate the Mandel Q parameter. This results in $Q \approx c^2 / ( 2(1+c)^2 \ln{(1+c)} ) - 1 |_{c = 36} = -0.87$, which is close to the simulation result.

\end{document}